\documentclass[usenatbib]{mn2e}
\newcommand{\LCDM}{$\Lambda$CDM~}
\newcommand{\CIAO}{\emph{CIAO}\ }
\newcommand{\XMM}{\emph{XMM-Newton}\ }
\newcommand{\Chandra}{\emph{Chandra}\ }
\newcommand{\MEKAL}{\textsc{MeKaL}\ }
\newcommand{\ROSAT}{\emph{ROSAT}\ }
\newcommand{\Zsol}{\ensuremath{\mathrm{Z_{\odot}}}}

\usepackage{subfigure,times,epsf, graphicx} 
\input{epsf}

\begin{document}
\title[The nature of the ghost cavity in the NGC~741 group]{The nature of the ghost cavity in the NGC~741 group}
\author[N.N. Jetha et al]{Nazirah N. Jetha$^{1,2}$\thanks{E-mail:nazirah.jetha@cea.fr}, Martin J. Hardcastle$^3$, Arif Babul$^4$, Ewan O'Sullivan$^{5}$, \newauthor Trevor J. Ponman$^{2}$, Somak 
Raychaudhury$^{2}$, Jan Vrtilek$^{5}$ \\$^{1}$Laboratoire AIM, CEA/DSM - CNRS - Universit\'{e} Paris Diderot, DAPNIA/Service d'Astrophysique, B\^{a}t. 709, CEA-Saclay, F-91191 Gif-sur-Yvette C\'{e}dex, France\\$^{2}$School of Physics and Astronomy, 
University of Birmingham, Edgbaston, Birmingham B15 2TT\\$^{3}$School of Physics, 
Astronomy and Mathematics, University of Hertfordshire, College Lane, Hatfield, Hertfordshire AL10 9AB\\$^{4}$Department of Physics and Astronomy, University of Victoria, Victoria, BC V8P 5C2, Canada \\$^{5}$Harvard-Smithsonian Center for 
Astrophysics, 60 Garden Street, Cambridge, MA 02138 }

\date{}
\maketitle
\label{firstpage}
\begin{abstract}
We discuss the effects of energy injection into the intra-group medium
of the group of galaxies associated with NGC~741.  The X-ray emission
reveals a large bubble, which in the absence of a currently bright
central radio source, may have been inflated by a previous cycle of
nuclear activity .  If the bubble is filled with a light, relativistic
fluid we calculate that in expanding, it has done more than sufficient
work to counteract the energy lost from the intra-group medium via
radiative cooling; the bubble can provide this energy as it expands
and rises.  Using upper limits on the flux density of the plasma
filling the bubble at 330 MHz and 1.4 GHz, we derive constraints on
its electron energy distribution and magnetic field strength. We show
that the data require the high-energy cut-off of the electron spectrum
to be very low compared to the cut-offs seen in more typical radio
sources if the fluid filling the bubble is a conventional relativistic
plasma.  This suggests that the fluid filling the bubble may not have
evolved by expansion or synchrotron losses consistent with a dead
radio source, leaving a puzzle as to what the origin of the bubble may
be.
\end{abstract}
\begin{keywords}
X-rays:galaxies:clusters - intergalactic medium - galaxies:active - galaxies:clusters:individual(NGC~741)
\end{keywords}

\section{Introduction}
\label{introduction}

It is well known that there is a lack of very cool gas in the centres
of galaxy groups and clusters (e.g. \citealt{2001A&A...365L.104P} and
\citealt{2002A&A...391..903S}), even in systems where the cooling time
is significantly shorter than the Hubble time.  Why this should be so
remains a mystery, and suggests that there must be some mechanism to
prevent catastrophic cooling.

If a bubble inflated by an active galactic nucleus (AGN) is over-pressured
with respect to the inter-galactic medium (IGM), then it will expand,
doing work on the IGM \citep[see for example][]{2006astro.ph..2566N}.
Depending on how significantly the bubble is overpressured, the expansion
may be trans or super-sonic, driving shocks into the IGM
(e.g. in Centaurus~A \citealt{2003ApJ...592..129K}), or the expansion
may be more subtle, doing gentle $PdV$ work on the gas as the bubble
expands (e.g. \citealt{2001ASPC..240..363B} and
\citealt{2002AAS...201.0304R}).  Once the bubble has expanded
and reached pressure equilibrium with the external medium, it will
begin to rise buoyantly.  As it rises, it may distort
\citep[e.g.][]{2001ApJ...554..261C}. The bubble will also do further
work on the IGM whilst it rises.  Such bubbles could remain intact
until long after the AGN has switched off, with the spectrum of the
bubble plasma steepening so that the emission at higher radio
frequencies drops to below detectable rates.

Observationally there is much evidence for this picture.  A number of
clusters show radio signatures of jets terminating in lobes that, in
turn, are coincident, in projection with X-ray surface brightness
depressions that, on occasion are bounded by weak shocks
(c.f. \citep[e.g.][]{2004ApJ...607..800B} and references therein).
The simplest explanation is that the inflating radio bubbles are
displacing the X-ray emitting plasma.

There are, however, cavities with no detectable radio emission (see
for example Abell~1795, \citealt{2002MNRAS.331..635E}; Abell~2597,
\citealt{2005ApJ...625..748C}; HCG~62,
\citealt{2004ApJ...607..800B}).  These so-called `ghost cavities' are
typically assumed to be features created in the more distant past and
whose radio emission has faded over time.  This interpretation finds
some support in the fact that in any given cluster, the ghost cavities
are typically found further away from the cluster centre than the
cavities with detectable radio emission (c.f. \citealt{2004ApJ...607..800B}).

Whether or not a cavity is associated with radio emission, of course
depends critically on the nature of the fluid that fills the bubbles.
If the bubbles are inflated by AGN, then it would be likely that the
fluid is a relativistic plasma.  However, the lack of associated radio
emission in ghost cavity systems implies that the plasma must be in
such a state that any emission is below detectable levels.  The lack
of observed radio emission places strong constraints on the nature of
the radiating particles and the field.

In this paper we investigate a centrally located ghost cavity in the galaxy group
associated with NGC~741.  This is an optically identified group
\citep{1998ApJ...496...39Z}, consisting of approximately 48 galaxies
within 1~Mpc, with a mean redshift of 0.019 and a velocity dispersion $\sigma =
458\pm 66~\mathrm{km\ s^{-1}}$.  The BGG, identified as NGC~741,
and its companion galaxy NGC~742 are separated by 45~arcsec (17~kpc),
with NGC~742 lying almost exactly to the E of NGC~741, together with a
much smaller third companion approximately 9~arcsec (3.4~kpc) to the
NE of NGC~741 \citep{1995A&A...297...28B}.
\citet{1998ApJ...496...73M} first reported a significant extended
X-ray halo in this group, extending to 19.2~arcmin from the centre of
NGC~741 (approximately 440~kpc in our adopted cosmology), with a
luminosity of $5.3\times 10^{41}~\mathrm{erg\ s^{-1}}$ (again in our
adopted cosmology).

The organization of the rest of the paper is as follows.  We present
the X-ray data and our analysis in Section~\ref{xray}.  In
Section~\ref{radio} we present the radio data and also comment on the
dynamical state of the system.  In Section~\ref{energetics} we discuss
the energetics involved and in Section~\ref{physcond} we place limits
on the nature of the fluid filling the cavity.  Our conclusions are
given in Section~\ref{conclusions}.  Throughout this paper, we use
$H_0=70~\mathrm{km\ s^{-1}\ Mpc^{-1}}$, with a \LCDM cosmology.
NGC~741 is at a redshift of 0.019, and 1~arcsec=0.38~kpc.  All errors
are quoted at the 1-$\sigma$ level.

\section{X-ray data}
\label{xray}

The \Chandra data of NGC~741 (ObsID 2223) were obtained from
the archive, and processed as part of the study of
\citet{2007MNRAS.376..193J}; the data were reprocessed using \CIAO~3.4
and {\sc caldb}~3.3.0 following the \CIAO threads online to apply new
calibration products and a new event file was created.  The data were
cleaned using a 3$\sigma$ clipping algorithm to remove times of high
background, leaving a 30.3~ks exposure.
	
In addition, we obtained and reduced the \XMM data (sequence number
0153030701) for the source (originally published in
\citealt{2005ApJ...622..187M}), using SAS version 7.0.0.  The original exposure times were
8.7~ks (MOS) and 7.0~ks (PN).  The data were filtered for flares
following the method described in
\citet{2007A&A...461...71P}, and vignetting corrections were applied
to the data following the procedure described in
\citet{2001A&A...365L..80A}.  After cleaning, we were left with good
time intervals of 5.8~ks (MOS1), 5.9~ks (MOS2) and 3.8~ks (PN).

\subsection{Spatial analysis}
\label{spatial}

\begin{figure*}
\scalebox{.3}{\includegraphics{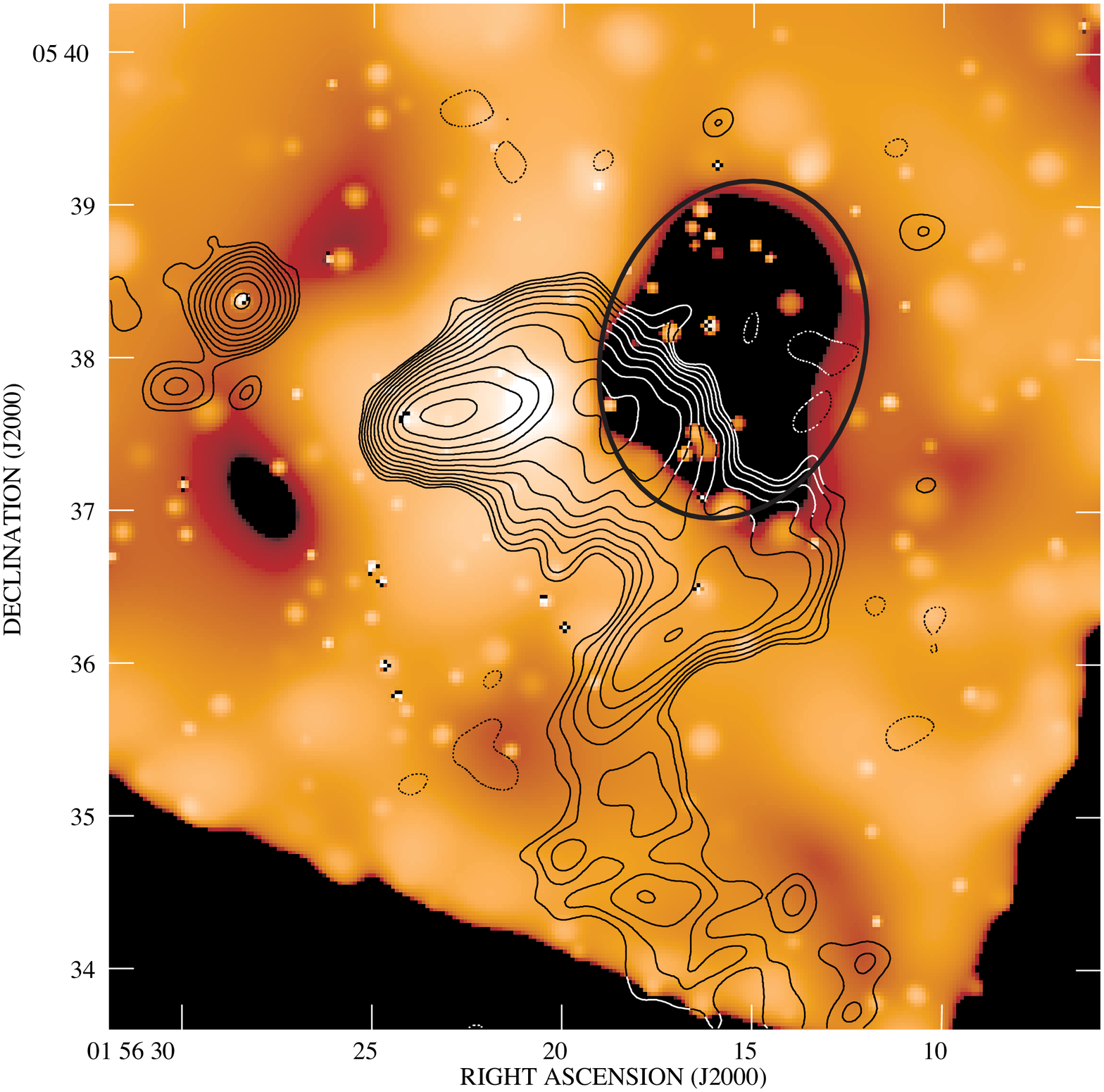}}
\scalebox{.4}{\includegraphics{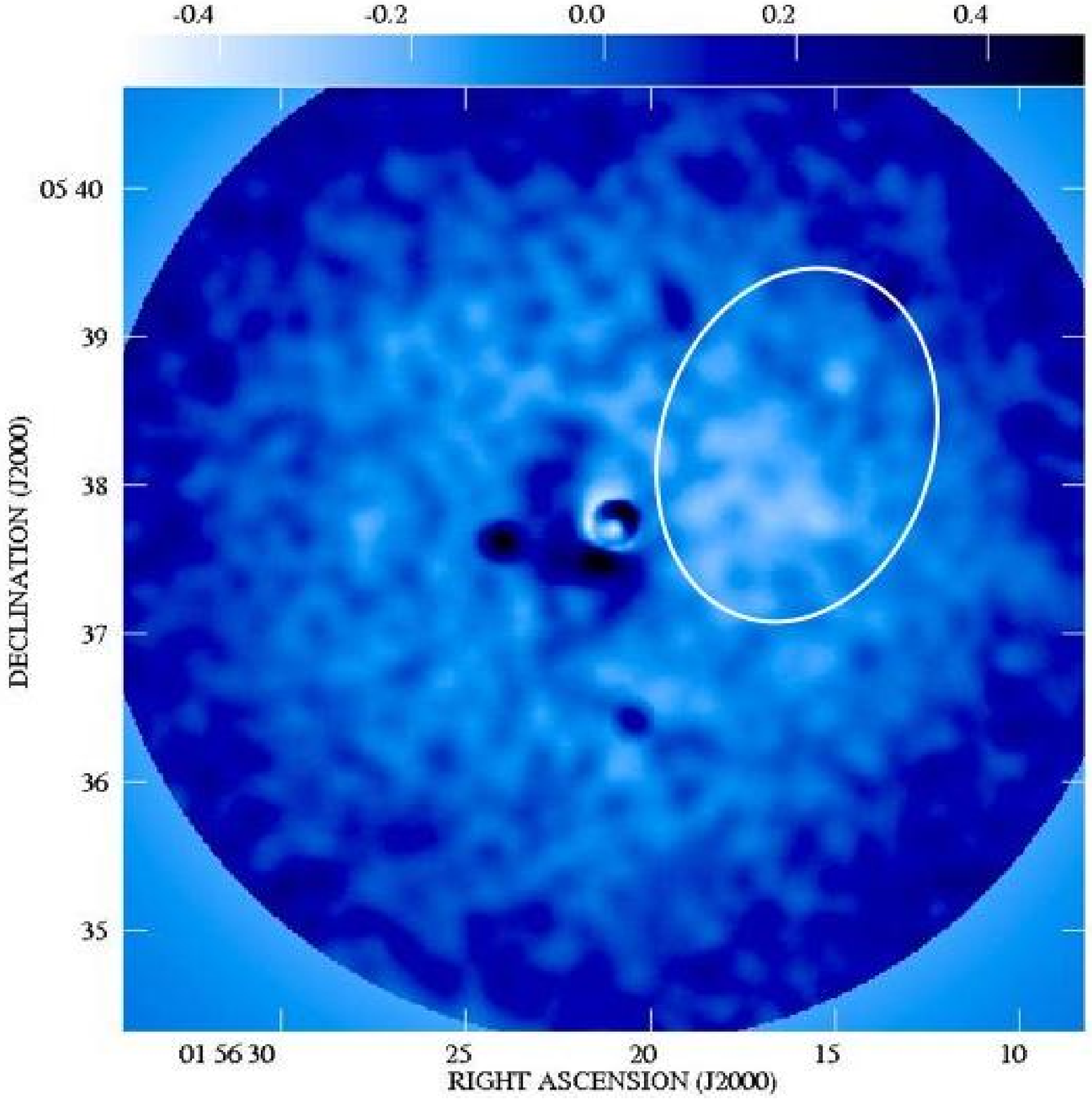}}
\caption{{Smoothed, exposure corrected \Chandra images of NGC~741 in
the energy range 0.5--5.0~keV as described in Section~\ref{spatial}.
The left panel shows the {\sc csmooth} image overlaid with 1.4~GHz VLA
contours, whilst the right panel shows the Gaussian smoothed residuals
to the 2-D fit.  In the left panel, the cavity described in Section~\ref{spatial} is
clearly visible, and is marked by the black ellipse. The cavity is
also clearly seen in the residual image.}}
\label{image1}
\end{figure*}

We smoothed the exposure corrected and binned \Chandra data in the
0.5--5.0~keV energy range using the task {\sc csmooth} with a minimum
smoothing signal to noise ratio of $3~\sigma$ (see Fig.~\ref{image1}
left panel).  This process revealed an elliptical cavity to the west
of the host galaxy (see also Vrtilek et al (in prep)).  Increasing the
minimum smoothing significance to 4~$\sigma$ resulted in a somewhat
smaller cavity.  We use an elliptical region to approximate the
boundaries of the cavity seen with 3~$\sigma$ smoothing; this region
has semi-major axis, $a_b\sim$19~kpc and semi-minor axis, $b_b
\sim$14.5~kpc, (marked by the elliptical region in Fig~\ref{image1}).
We assume a prolate ellipsoid so that the third axis $c_b=b_b$.

\begin{figure*}
\subfigure{\scalebox{.4}{\includegraphics{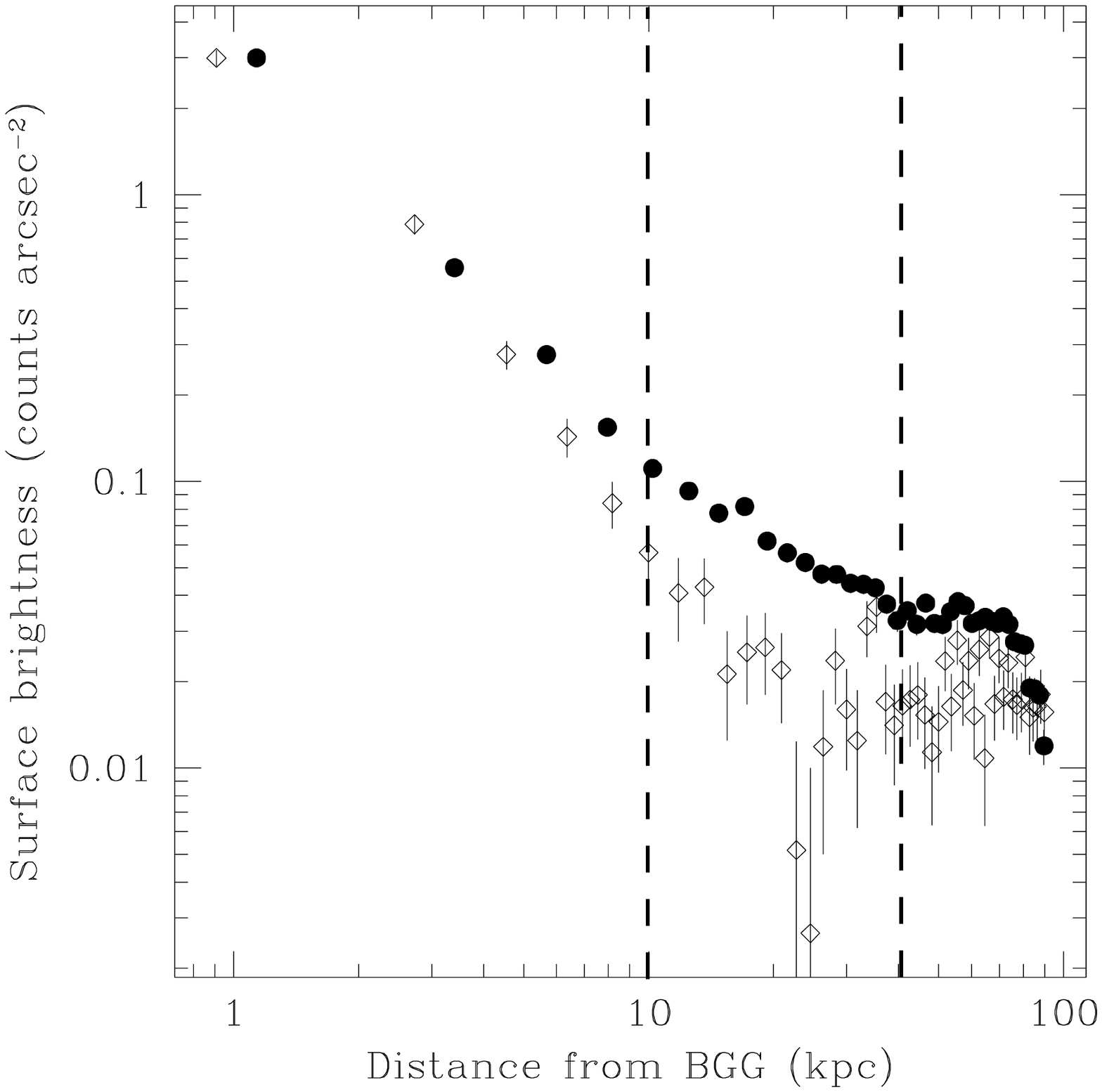}}}
\subfigure{\scalebox{.4}{\includegraphics{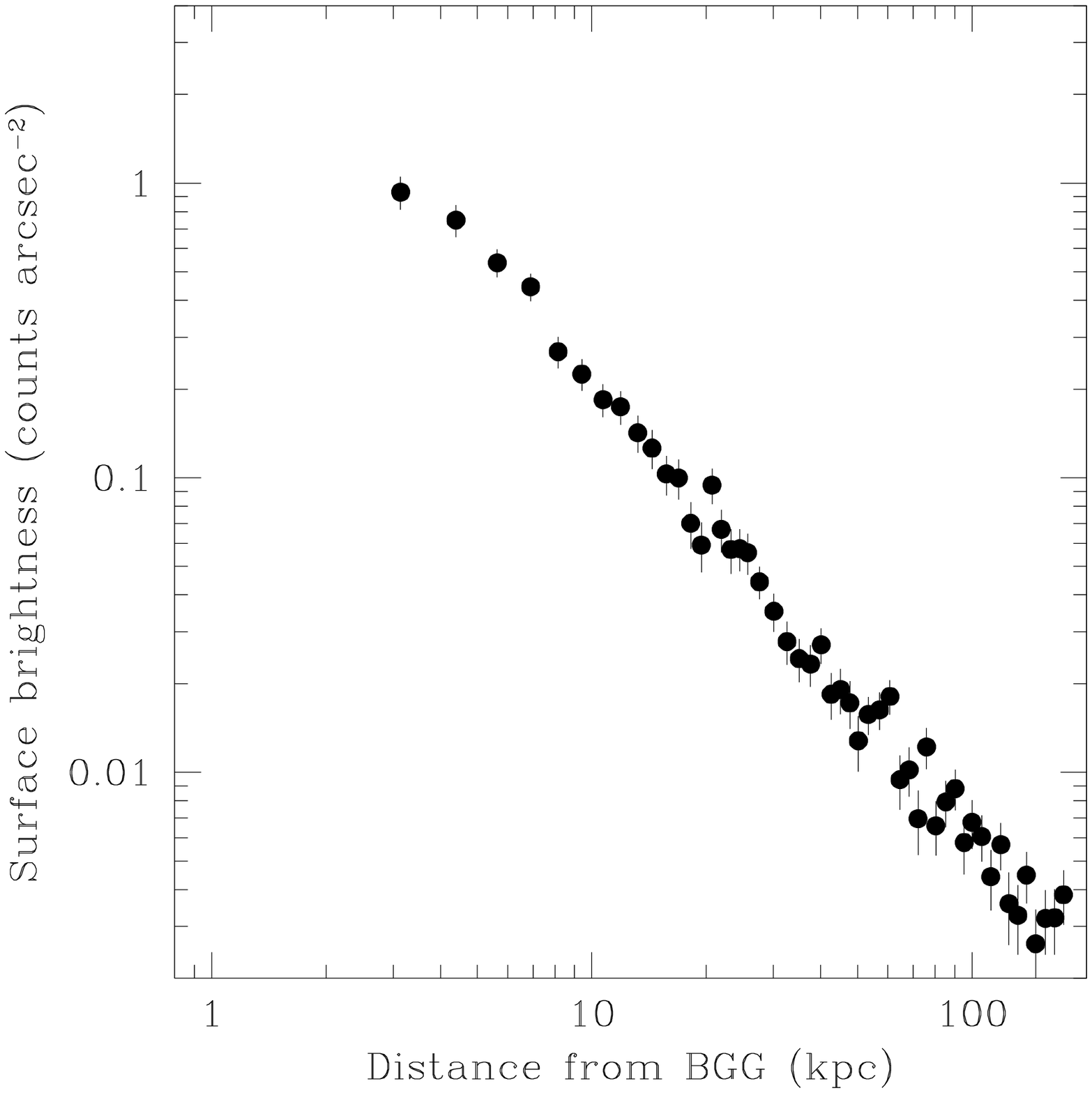}}}
\caption{\Chandra (left) and \XMM (right) surface brightness profiles
for the NGC~741 group.  For the \Chandra data, profiles were taken in
wedges in the direction of the X-ray cavity (open points) and the
undisturbed gas (solid points).  There is at least a 3~\(\sigma\) drop
in surface brightness in the region between the dashed lines.  The
\XMM profile is azimuthally averaged and uses data from the M1, M2,
and PN instruments of \XMM.  The profile was fitted with a beta-model
and this fit was used in Section~\ref{explainmorph} to model the large
scale emission of the group.}
\label{sbprofC}
\end{figure*}

To check that this cavity is real and not just an artefact of adaptive
smoothing we firstly created an exposure corrected image in the
0.5--5.0~keV energy range as above and smoothed it with a fixed
Gaussian kernel of FWHM 5~arcsec.  We then used the {\it Sherpa}
fitting program to fit a single 2-D $\beta$-model to the \Chandra data
to model the large-scale emission.  This single-component fit is
intended to model the large-scale group emission, allowing it to be
subtracted from the image and revealing structures in the core. As a
single-component model can only approximate the true surface
brightness distribution, we do not use this model to derive physical
parameters.  The model centre is slightly off-centre due to the
structure in the centre of the group.  We then created an image of the
residuals of the fit, which is shown in Fig~\ref{image1} (right
panel).  The cavity can be clearly seen, and we overlay the same
ellipse as in the left panel to guide the eye.  We further created a
surface brightness profile in the 0.5--5.0~keV energy range which was
centred on NGC~741, and covered, in azimuthal bins, the region
occupied by the cavity, together with a comparison surface brightness
profile in the same energy range for the entire system excluding the
region we used to extract a profile of the cavity.  The opening angles
of the wedges used for the two profiles were 80 and 280~degrees
respectively, and a local background was used for background
subtraction.  The resulting profiles are shown in Fig.~\ref{sbprofC}
(left panel) overlaid on each other.  There is a clear 3~$\sigma$
surface brightness drop which begins at approximately 10~kpc from
NGC~741 and continues to approximately 40~kpc, implying that that the
cavity seen in the images is real. At the centre of the cavity,
the ratio, $\mathbf{R_{\mathrm{SB}}}$ of the surface brightness of the cavity,
to that of the undisturbed gas, is approximately
$\left(0.4\pm0.1\right)$.  Compared to other systems, where the
surface brightness contrasts are of the order of 20--30\% of the
surrounding surface brightness (\citealt{2004ApJ...607..800B} and
\citealt{2006ApJ...652..216R}), the bubbles here are of a similar
surface brightness contrast.

\begin{figure}
\scalebox{.4}{\includegraphics[angle=270]{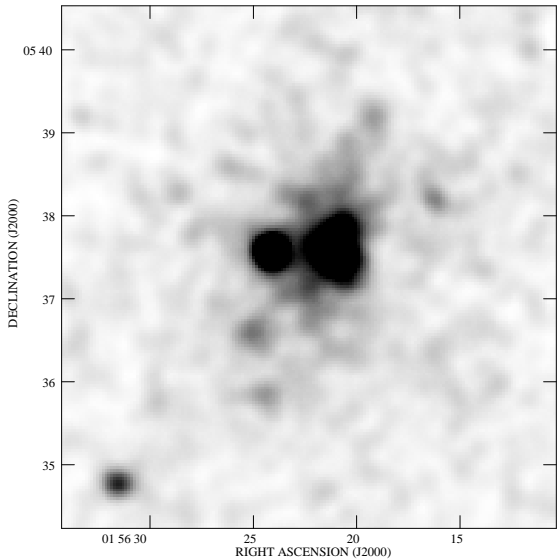}}
\caption{\XMM mosaiced image, using the M1, M2 and PN cameras of the group associated with NGC~741.  The data have been smoothed with a Gaussian beam of FWHM 6~arcsec.}
\label{xmm_smth}
\end{figure}

For completeness, we generated a surface brightness profile in the
0.3--2.0~keV energy range from the \XMM data (shown in
Fig~\ref{sbprofC} right panel).  We use this profile to provide a
model for the surface brightness for use in Section~\ref{explainmorph}.
We used the \XMM rather than the
\Chandra data for this since the \XMM data extends out to larger
radii, providing a better model for the large scale emission.  The
profile was fitted by a $\beta$~-model with $\beta = 0.428\pm0.005$
and core radius, ($r_c = 2.1\pm0.2$)~kpc. We also smoothed the \XMM
data with a Gaussian beam of FWHM 6~arcsec (shown in
Fig.~\ref{xmm_smth}) to see if the cavity was detectable in the data.
Whilst the cavity is not visible in the \XMM data due to PSF
blurring and the shorter exposure time from the deficit seen in the
\Chandra data, we would expect the surface brightness as viewed by \XMM to drop from
0.35~$\mathrm{counts\ arcsec^{-2}}$to 0.14~$\mathrm{counts\
arcsec^{-2}}$.  In the \XMM data, we observe a surface brightness of
$0.2\pm0.5~\mathrm{counts\ arcsec^{-2}}$ in the region defined by the
cavity, consistent at the 1-$\sigma$ level with the cavity seen in the
\Chandra data.

\subsection{Spectral analysis}
\label{spectral}

From the \XMM data, we extracted a global spectrum from a source
region of 8.8~arcmin centred on the BGG. The
source region was chosen as we detect emission at 3-$\sigma$
significance out to this radius.  The X-ray background is subtracted
using the background files of \citet{2003A&A...409..395R} and we use
the double subtraction method of \citet{2002A&A...390...27A} to take
into account both the X-ray and the particle backgrounds correctly.
The spectrum was fitted with a simple absorbed \MEKAL model, the
details of which are given in Table~\ref{mekal}.  We find a
temperature of $1.7\pm0.1$~keV, with metal abundances fixed to
$0.3$~\Zsol.  The fit has $\chi^2=147$ for 115 degrees of freedom.  We
find a luminosity for the group of $6.6_{-0.9}^{+0.4}\times
10^{41}~\mathrm{erg\ s^{-1}}$ which agrees well with the
\citet{1998ApJ...496...73M} \ROSAT result when differing cosmologies
are taken into account.  Our value for temperature agrees well with
that of \citet{2005ApJ...622..187M}, who use the same \XMM data.

\begin{table}
\caption{Details of the \MEKAL model fitted to the \XMM data for all
three instruments (M1, M2 and PN).  The M1 and M2 temperatures were
tied together, and metalicities were fixed to 0.3~\Zsol.  The hydrogen
column density was initially fixed to the Galactic value, but then
allowed to vary.  This did not produce a significant difference in the
fits. We fitted an area out to 8.8~arcmin, as described in the text.}
\label{mekal}
\begin{tabular}{lll}
\hline
Component&Parameter&\\
\hline
Absorption&$N_H(\times 10^{22} \mathrm{cm ^{-2}}) (Galactic)$&0.044\\
Absorption&$N_H(\times 10^{22} \mathrm{cm ^{-2}}) (Fitted)$&$0.042\pm0.02$\\
M1              &                                 & \\
\MEKAL& $kT$ (keV)&$1.7\pm 0.1$\\
              &Norm&$(1.69^{+0.07}_{-0.1})\times 10^{-3}$\\
\hline 
M2 && \\
\MEKAL&Norm&$(1.69^{+0.09}_{-0.07})\times 10^{-3}$\\ \hline
PN && \\
\MEKAL& $kT$ (keV)&$1.6\pm 0.1$\\
              &Norm&$(1.06^{+0.03}_{-0.08})\times 10^{-3}$\\ \hline
              &$L_X (10^{41} \mathrm{erg\ s^{-1}})$&$6.6_{-0.9}^{+0.4}$\\
&$\chi^{2}$(d.o.f.)&147(115)\\
\hline
\end{tabular}
\end{table}

\begin{figure*}
\subfigure{\scalebox{.3}{\includegraphics{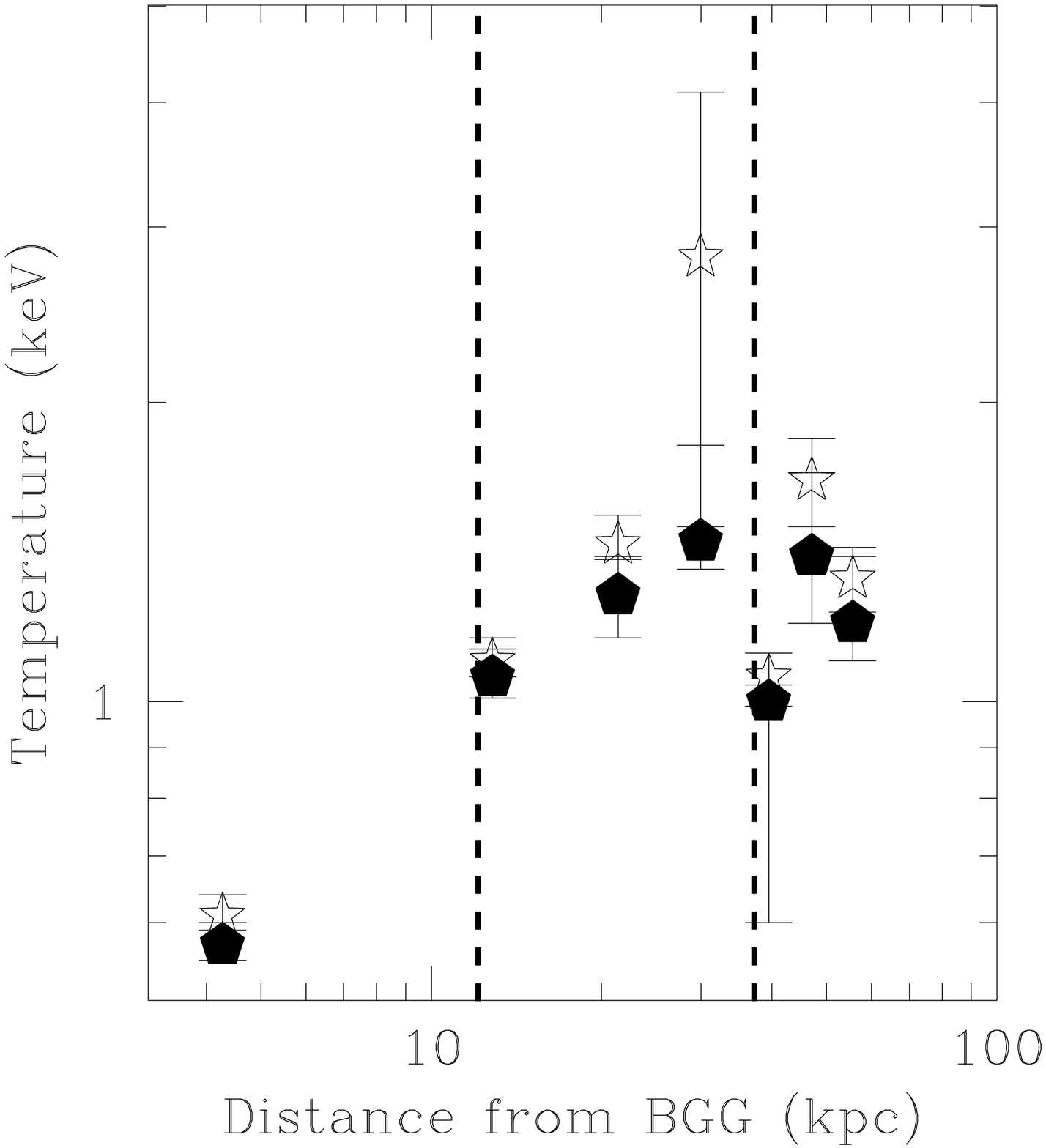}}}
\subfigure{\scalebox{.3}{\includegraphics{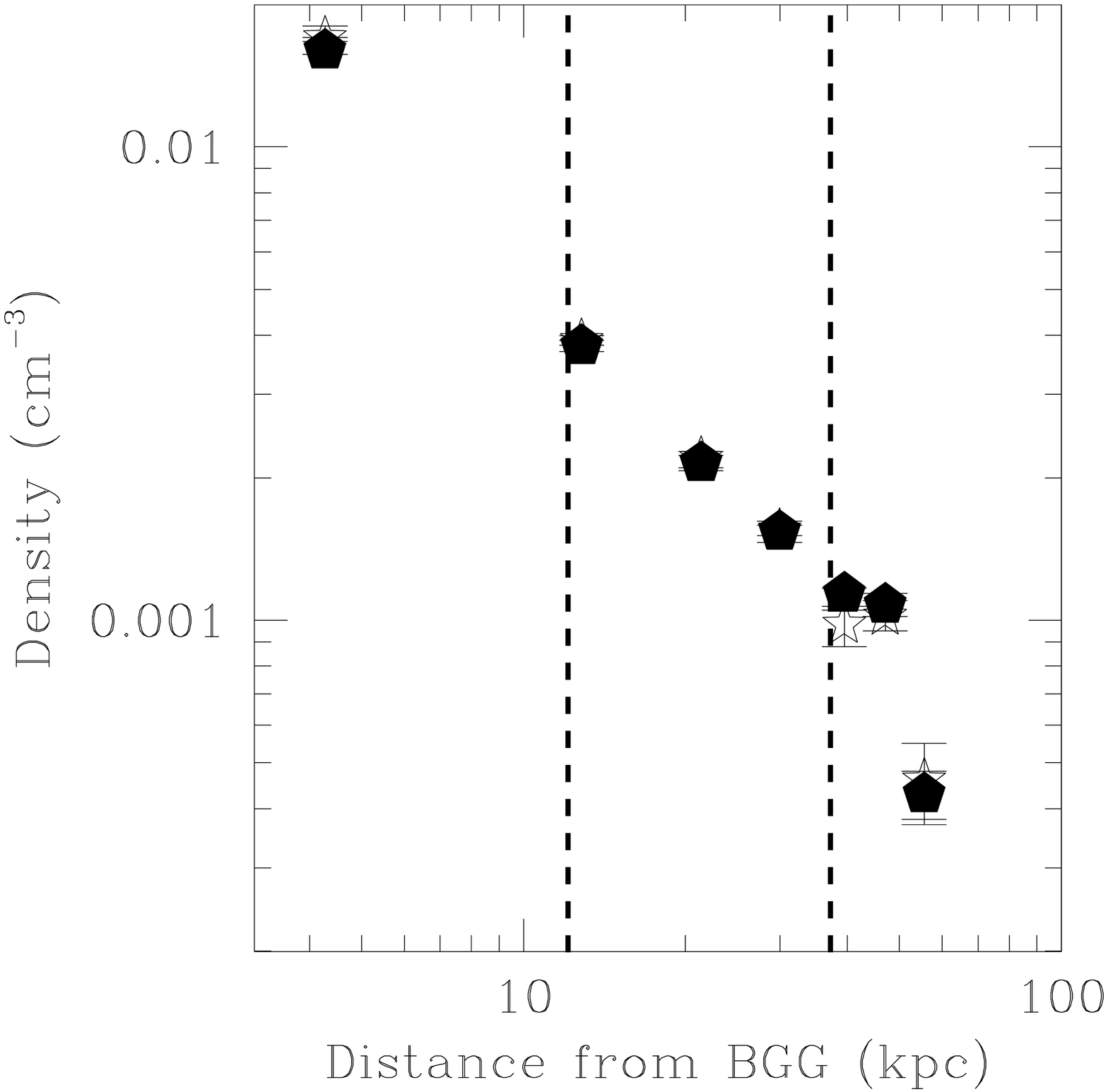}}}\\
\subfigure{\scalebox{.3}{\includegraphics{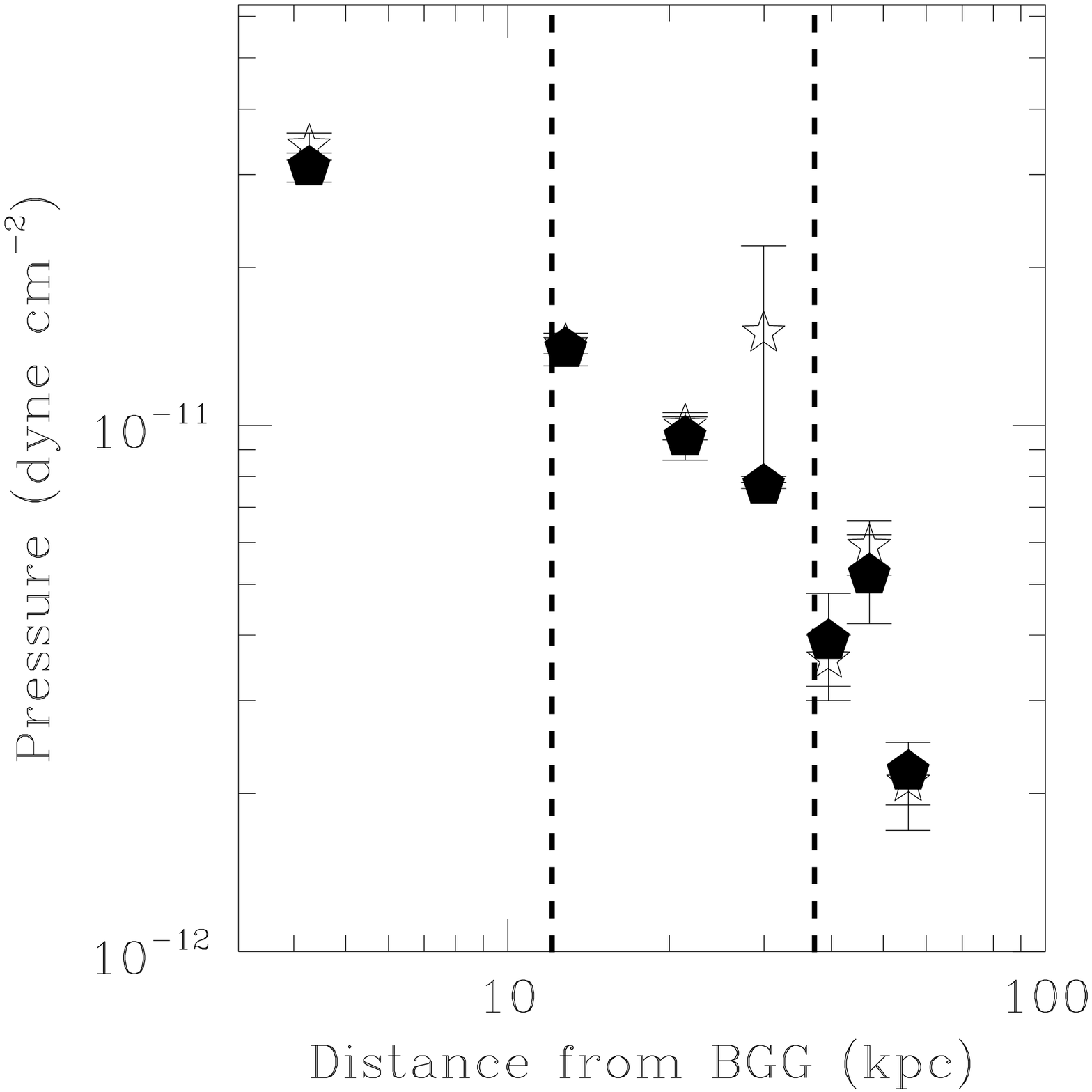}}}
\subfigure{\scalebox{.3}{\includegraphics{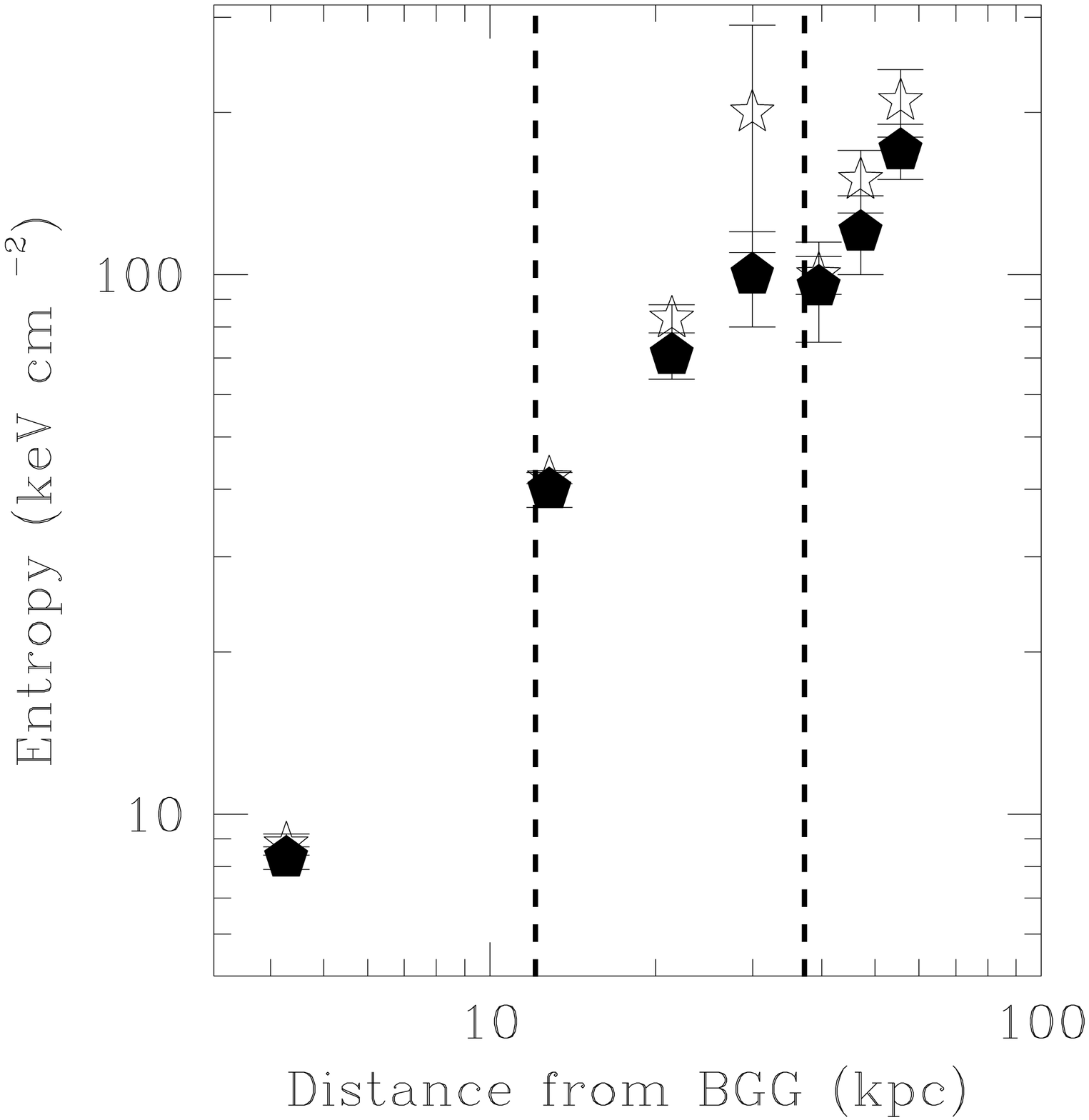}}}
\caption{The temperature, density, pressure and entropy profiles of the hot gas obtained from the \Chandra data.  The open points show the data 
for the entire group, whilst the solid points show the profiles for
the group with a region containing the bubble removed.  The vertical
dotted lines show the extent of the cavity as described in
Section~\ref{spatial} determined by the smoothed image, assuming that
the centre of the cavity is located at 30~kpc from NGC~741 (see Section~\ref{explainmorph}).}
\label{fig4}
\end{figure*}

We next generated azimuthally averaged temperature profiles from both
the \Chandra and \XMM data.  The profiles were again centred on the
BGG, and an ellipse defining the region containing the bubble was
excluded from the \Chandra data as shown in fig.~\ref{image1}.  For
the \Chandra data, annular regions were defined out to 70~kpc
(corresponding to 3.4~arcmin), and a blank sky background of the same
size and placement on the chip as the source region was used for
background subtraction.  The background files were processed in the
same way as the data (following the instructions in the {\it CIAO}
threads) and normalized such that both background files and the data
had the same flux in the 8--10~keV energy band.  Similarly, for the
\XMM data, annuli were defined out to 8.8~arcmin and background
regions as described earlier were used.  We then followed the method
described in \citet{Jethaa} for the \Chandra and \XMM data separately,
fitting an absorbed \MEKAL model to each annulus to deproject the
spectra, to produce the deprojected temperature profiles shown by the
open points in Fig.~\ref{fig4}a for \Chandra.  We do not show the \XMM
points here, but the \Chandra and \XMM data are in good agreement in
the regions of overlap.  The normalisations of the fitted \MEKAL
models were used to obtain the electron density profile
($n_e\left(r\right)$), since:
\begin{equation}
n_e = \left\{\frac{4\pi\left[D_A\left(1+z\right)\right]^2
N_{mek}\times 1.17}{1\times10^{-14}V}\right\}^{\frac{1}{2}},
\label{densityeq} \end{equation} where $D_A$ is the angular size
distance to the source, $z$ is the redshift of the source, $N_{mek}$
is the normalization of the \MEKAL component fitted to the annulus,
and $V$ is the volume of the annulus from where the spectrum was
extracted.  

In Fig.~\ref{fig4}, we overplot two different deprojections.  The open
points show the deprojection when we disregard the presence of the
cavity and treat the IGM as azimuthally symmetric.  The filled points
show the deprojection when we take the presence of the cavity into
account and exclude a region corresponding to the cavity from
deprojection.  As the volume excluded due to the cavity can be a
significant fraction of the volume of the shell in this case, this
must be taken into account when doing the deprojection.  We dealt with
the excluded volume in calculating $V$ in Eqn.~\ref{densityeq} by
scaling the volume of an individual shell according to the fractional
area of the annulus that was used to extract the spectrum.  Thus, for
an annulus towards the centre of the bubble, the volume excluded was
close to the elliptical volume of the bubble that contributed to the
shell, whilst for an annulus which only contained the edge of the
bubble, the excluded volume may have overestimated the contribution of
the bubble to the shell.

The temperature and density profiles are shown in the top two panels
of Fig~\ref{fig4}.  We further
derived the pressure profile of the data (shown in the bottom left
panel of Fig.~\ref{fig4}) using \begin{equation} P(r)=2.96 \times10^{-9}n(r)T(r),
\label{pressure}\end{equation} where the numerical factor is used to
convert the temperature and density to cgs units, $n\left (r\right)$
is the electron density profile, and $T\left(r\right)$ is the
temperature profile.  To ensure that our findings are robust, we redid
the deprojection with different regions for the spectral extraction.
We found that our profiles were generally consistent with each other,
regardless of the extraction regions chosen for the deprojection.

The entropy profile shown in the bottom right panel of Fig~\ref{fig4}
shows marginal signs of disturbances caused by the bubbles in the gas
surrounding them.  This suggests that the bubbles may be doing
work on the IGM.  Comparing our profiles to those of Mahdavi et al, we find
that our profiles and those of \citet{2005ApJ...622..187M} agree well
within the core of the group and at large radii.  However, in the
region between 20--60~kpc, \citet{2005ApJ...622..187M} find a slightly
lower entropy (but consistent within the errors) than what is found
here.  This discrepancy arises because the structure in this region is
complex and may be significant enough to affect the results of any
deprojection; thus the different method of deprojection used in
\citet{2005ApJ...622..187M} will produce slightly different results.
Our method of radial deprojection may well average over real
asymmetric structure, but comparison of profiles obtained via
different methods suggests that this does not change the final values
too drastically.
 
\subsection{Actual location of the cavity}
\label{explainmorph}

In Section~\ref{spatial} we calculate that the ratio of the surface brightness
at the centre of the cavity compared to the undisturbed surface
brightness, which we termed $\mathbf{R_{\mathrm{SB}}}$ in  Section~\ref{spatial}, is $0.4\pm0.1$.
We can use this ratio, using a method similar to that of
\citet{2007ApJ...659.1153W}, to estimate the three-dimensional location of
the cavity within the X-ray emitting plasma.  We assume that the
undisturbed IGM of the NGC~741 can be described by the azimuthally
averaged fit to the \XMM data, using the $\beta$-model as fitted in
Section~\ref{spatial}.  The radial distance of the bubble from the
centre of this spherically symmetric density profile is 25~kpc as
measured from the X-ray maps (Fig~\ref{image1}).  The unknown is the
distance of the cavity centre along the line of sight, which we denote
$z$ (taking $z=0$ to be the mid-plane of the spherically symmetric
group).  For a given $z$ the measured density profile allows us to
calculate the surface brightness drop produced by a cavity of the
observed dimensions, at the observed radial distance.

From the data, $\mathbf{R_{\mathrm{SB}}}=0.4\pm0.1$ (Section~\ref{spatial}); if the
cavity were at $z=0$, then we calculate that $\mathbf{R_{\mathrm{SB}}} = 0.3$, which is
within the uncertainty on our measured value. Thus it is
possible that the cavity is in the $z=0$ plane.  We find that a cavity
placed 14~kpc away from the $z=0$ plane reproduces a ratio of
0.4. Given the errors on the decrease in surface brightness, we can
calculate an error in the location of the cavity; the bubble could be
in the $z=0$ plane, or as far away as 22~kpc from the $z=0$ plane.
Combining this with the measured radial distance of 25~kpc we find
that the cavity is located at a physical distance of (29$\pm$4)~kpc
from the centre of the NGC~741 group. Taking this uncertainty into
account, the ambient temperature and pressure of the cavity are in the
range 1--2~keV and 3--10.4$\times 10^{-12}~\mathrm{dyne~cm^{-2}}$
respectively  (taken from the deprojection in which the cavity was
  excluded: Fig~\ref{fig4}). This range of temperatures and pressures has
implications for both the energy imparted to the IGM and the time-scale
over which the energy can be imparted, which we discuss in
Section~\ref{energetics}.

\section{Radio sources in the system}
\label{radio}
NGC~741/2 has been known to be associated with a relatively bright
radio source (4C 05.10) since early work by \citet{1980MNRAS.192..595J}
and \citet{1985ApJ...291...32B}. Radio images showed two bright peaks
close to the centres of the two galaxies surrounded by more diffuse
emission.   \citet{1985ApJ...291...32B} argued that all the radio
emission, including the bright peak close to the nucleus of NGC~742,
was associated with NGC~741: on this interpretation NGC~741 would have
been something akin to a classical double source and the compact
source near NGC~742 would have been its hotspot. It is probably
because of this interpretation that NGC~741 is generally thought of as
being the radio galaxy of the system. \citet{1994ApJ...436...67V} used
VLA observations to show that the component close to NGC~742 was truly
compact and well aligned with the centre of the galaxy, and proposed
instead that the diffuse radio emission was a head-tail or
narrow-angle tail (NAT) source associated with NGC~742.

\begin{figure*}
\scalebox{.8}{\epsfbox{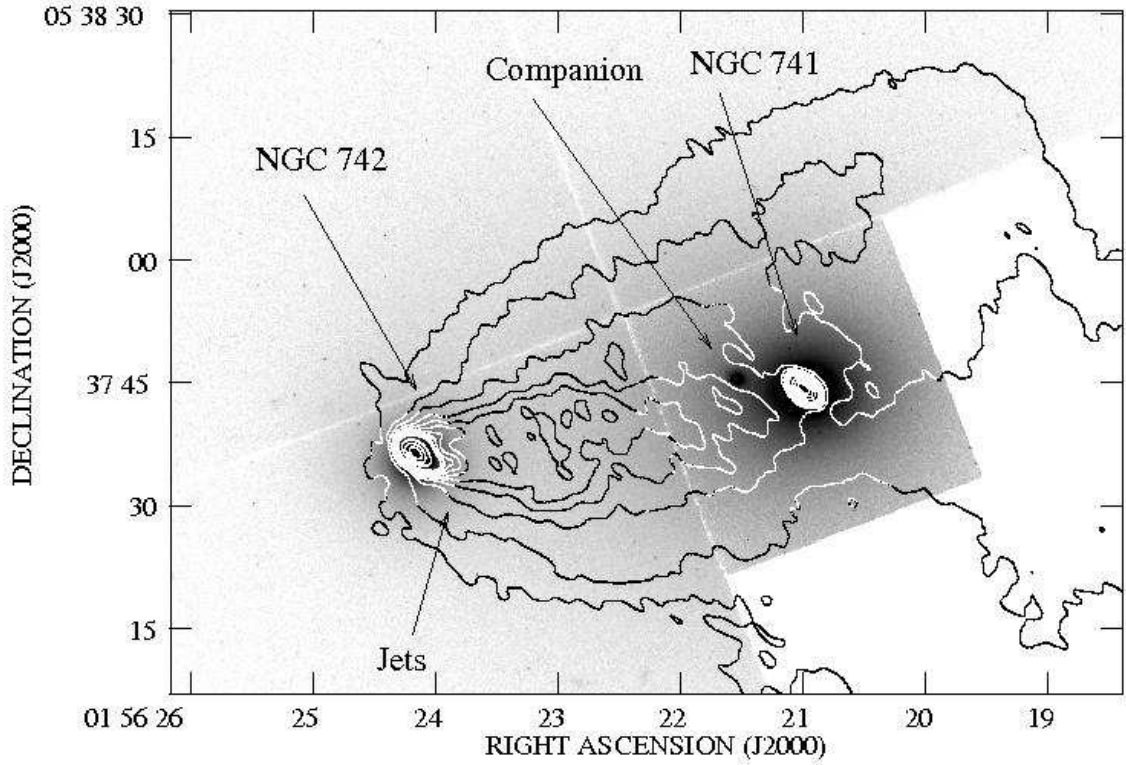}}
\caption{Radio and optical emission around NGC~741/2. The greyscale shows the {\it HST} F814W image described in the text, with grey levels 
chosen to emphasise the common stellar envelope of the
system. Contours are from our 1.5-GHz radio image with resolution $3.1
\times 1.7 $~arcsec (major and minor FWHM of elliptical restoring
Gaussian) at $0.5 \times (1,2,3\dots
10,15,20\dots)$~mJy~beam$^{-1}$. The three optical galaxies and the
bent jets of NGC~742 are labeled.}
\label{overlay}
\end{figure*}

To allow us to make the most definite statements possible about the
nature of the radio emission, we have acquired multi-frequency VLA
observations (listed in Table \ref{vlaobs}) of the system from the
archive and reduced them in the standard manner in {\sc AIPS}. For
display purposes we combine these with a {\it Hubble Space Telescope}
images taken with the WFPC2 instrument on 1997 Jun 20, obtained from
the public archive. In the images obtained from the data (e.g.\
Fig.~\ref{overlay}), the two compact sources seen in earlier
observations are well aligned with the nuclei of the two
galaxies. However, twin jets are clearly visible emerging from the
nucleus of NGC~742 and bending westwards to merge into the
larger-scale extended emission, while there is no small-scale
extension of any kind associated with the nucleus of NGC~741. (A third
bright radio source in the field, a small triple whose central
component has a position of RA 01$^h$ 56$^m$ 28.5$^s$ Dec. $+05^\circ$
$38'$ $24''$, is coincident with a very faint point source in the {\it
HST} image and is most likely a background quasar.) The lack of any
extension in the NGC~741 nuclear source is consistent with the results
of \citet{2000AJ....120.2950X}, who show that the source is unresolved even on
parsec scales. In the VLA data the compact components have flat radio
spectra between 1.5 and 4.9~GHz, indicating that there is
self-absorbed parsec-scale structure in both.

We therefore agree with \citet{1994ApJ...436...67V} that it is
entirely plausible that the extended radio emission in the NGC~741/2
system comes only from a narrow-angle tail source associated with
NGC~742.  The 400~km~s$^{-1}$ difference in the radial velocities of
the two galaxies \citep{2004ApJ...607..202M} is consistent with a
narrow-angle tail interpretation of the radio emission, since the much
more massive NGC~741 is likely to be close to the centre of mass of
the group (its radial velocity is the same as the mean of the 48
galaxies within 1~Mpc -- 5570~$\mathrm{km\ s^{-1}}$). The common
stellar envelope of the two systems (clearly visible in Fig.~
\ref{overlay}) implies that they are close enough to be interacting
tidally, but they need not be as close as they appear in projection
(18~kpc): if NGC~742 were displaced by a few tens of kpc along the
line of sight with respect to NGC~741 it would help to explain the
lack of correspondence between the X-ray emission (dominated by gas on
smaller scales) and the observable radio emission, and in particular
the absence of any X-ray cavities associated with the radio source.
NGC~741 clearly hosts some form of active galaxy, but has no evidence
for current large-scale jets. 

We further argue that the bubble visible in the system is associated
with a previous outburst of NGC~741 rather than with NGC~742 for
several reasons.  Firstly, NGC~742 already shows evidence of old
emission in the form of a long tail; if the cavity had been caused by
NGC~742, then the bubble must be older than the current large scale
emission.  However, the projected distance from the core of NGC~742 to the
end of the tail is 160~kpc, while the bubble is located only 45~kpc
away.  If the bubble is from a previous outburst of NGC~742, then we
would have expected it to have travelled further than the observed
tail from the current outburst.  Secondly, to have got to its current
position relative to NGC~742, the bubble would have had to have
crossed the centre of the group against the pressure gradient.
Thirdly, if the radio emission from NGC~742 were removed from the
picture, NGC~741 would look much more similar to some other
well-studied `ghost cavity' systems like NGC~4636 \citep
{2002ApJ...567L.115J} or HCG 62 \citep{2006PASJ...58..719M}, with a
weak central radio source bearing no obvious relation to the
observable cavity.  Therefore, in the remainder of the paper we adopt
the hypothesis that the bubble was caused by a previous outflow of
NGC~741 and that the location of the large scale emission of NGC~742
is entirely coincidental.

\begin{table}
\caption{Radio observations used in our analysis}
\label{vlaobs}
\begin{tabular}{llrrr}
\hline
VLA obs.&Date&Configuration&Frequency&Time on source\\
ID&&&(GHz)&(min)\\
\hline
AB920&1999 Jul 18&A&1.5&50\\
AB593&1991 Dec 16&B&1.5&10\\
&&C&4.9&26\\
AH276&1988 May 19&C&1.5&30\\
&&C&4.9&76\\
AS827&2005 May 07&B&373\\ \hline
\end{tabular}
\end{table}

\section{Energy imparted to the IGM}
\label{energetics}

In Section~\ref{introduction}, we discussed the possibility that bubbles
will do mechanical work on the IGM by expanding and rising, preventing
catastrophic cooling.  Here, we calculate the amount of energy
available to do this.  The physics of AGN-inflated bubbles is
discussed in length in the literature
(e.g. \citealt{2001ASPC..240..363B}, \citealt{2002AAS...201.0304R} and
\citealt{2006astro.ph..2566N}), and in all cases, the initial
overpressure factor is unknown.  We now briefly investigate the total
amount of energy available based on whether the bubble was inflated
gently (as discussed in \citet{2006MNRAS.372...21A},
where the bubble is overpressured by a factor of 2 with respect to the
IGM), or if the bubble was injected more violently, say if the
overpressure factor was 10, as seen in some young low-power radio
sources \citep{2003ApJ...592..129K}.

Assuming that the bubble lies 29~kpc from the centre of NGC~741 and
that the ambient pressure is in the range $3-10\times
10^{-12}~\mathrm{dyne\ cm^{-2}}$ (from Fig.~\ref{fig4}c), we
calculate, (using eqn. 18 of \citet{2006astro.ph..2566N} and assuming
that $\Gamma=4/3$), that if the bubble is overpressured by a factor of
2, then the energy imparted to the IGM is
$E_{IGM_{2}}=\left(0.5-1.7\right)\times 10^{58}$~erg.  If however, the
bubble is overpressured by a factor of 10, then
$E_{IGM_{10}}=\left(1-4\right)\times 10^{58}$~erg.

To compare this with the radiative losses of the IGM, i.e. the X-ray
unabsorbed luminosity within the region that the bubbles could have
affected, a characteristic time-scale needs to be used.  This time-scale
over which this is done is also important, since if the energy is
injected too slowly, it will be insufficient to counteract cooling,
and if the energy is released too quickly, it would cause a spike in
the temperature of the IGM.

Following \citet{2004ApJ...607..800B}, we define three different
time-scales, and thus potential ages for the bubble.  The first ($t_c$)
is based on the time it would take for the bubble to rise to its
current location from the BGG at the group's sound speed, if the
bubble had been inflated by a powerful jet close to the radio core
(see for example the simulations of \citealt{2004MNRAS.348.1105O}).
The second estimate ($t_b$) is based on the assumption that the bubble
is again inflated close to the radio core but rises buoyantly rather
than being propelled due to its production mechanism.  The third
method is to calculate how long it would take to refill the volume
displaced by the bubble (\citealt{Hydra} and
\citealt{2002ApJ...568..163N}), which is denoted $t_r$.  We use
eqns.~2--4 from \citet{2004ApJ...607..800B} to calculate time-scales
based on the sound-speed ($t_c$), the buoyant rising time of the
bubble ($t_b$) and the refilling time for the bubble ($t_r$)
respectively.  The time-scale and powers calculated are shown in
Table~\ref{power} and the unabsorbed luminosity of the group within 29~kpc is
0.36$\pm0.02 \times 10^{42}~\mathrm{erg\ s^{-1}}$.  Thus, it is clear
that over the lifetime of the bubble, and for the foreseeable future,
energy input from the bubble will dominate over radiative losses (by
1--2 orders of magnitude), providing more than sufficient energy to
counteract cooling losses, even if heating is not 100\% efficient.

\begin{table*}
\caption{Energy injection rates based on different methods of bubble motion and bubble formation.}
\label{power}
\begin{tabular}{lcccccccc}
\hline
$x$ & $E_{\mathrm{IGM}}$ & $t_c$        & $t_b$    & $t_r$	   & $P_{t_c}$ & $P_{t_b}$ & $P_{t_r}$ & $L_X$ \\
       &$10^{58}$~erg &$10^7$yr&$10^7$yr&$10^7$yr&$10^{42}~\mathrm{erg\ s^{-1}}$&$10^{42}~\mathrm{erg\ s^{-1}}$&$10^{42}~\mathrm{erg\ 
s^{-1}}$&$10^{42}~\mathrm{erg\ s^{-1}}$\\
\hline
2   &0.5--1.7&  4.5 & 8.6 & 6.7 & 4--10 & 2--6 & 4--20 & 0.36 $\pm$0.02\\
10  & 1--4   &  4.5 & 8.6 & 6.7 & 7--30 & 4--15 & 9--30 & 0.36$\pm$0.02\\
\hline
\end{tabular}
\begin{minipage}{16cm}
$t_c$, $P_c$ are the time-scales and powers derived assuming that the
bubble was inflated by a powerful jet and rose to its current position
at the sound speed of the group.\\ $t_b$, $P_b$ are the time-scales and
powers derived assuming that the bubble was inflated by a more gentle
jet and rose to its current position buoyantly.\\ $t_r$, $P_r$ are the
time-scales and powers derived from refilling arguments.
\end{minipage}
\end{table*}

\section{Physical conditions in the bubble}
\label{physcond}
If the observed cavities in the X-ray emission correspond to bubbles
injected by an AGN outburst, then we can use the X-ray and radio data
available to constrain the nature of the plasma filling the bubbles.
Our starting assumption is that the pressure in the bubble must be at
least equal to the external thermal pressure, while it cannot be very
much greater than that pressure to avoid driving a shock into the
external medium. In addition, the bubble must be light compared to the
external medium: it is hard to see how an over-dense `bubble' could
have reached its present distance from the nucleus of NGC 741.

\subsection{Content of the bubbles}
There are then two basic possibilities for the nature of the bubble
material. The first is that it is very hot (i.e. $T_{\rm b} \gg T_{\rm
ext}$), low-density material (for example
\citealt{2003ApJ...596..159M}and \citealt{2007arXiv0706.2768M}).  The
second, and in our view more {\it a priori} likely, possibility for
the nature of the bubble material is that it consists wholly or partly
of a relativistic plasma, containing at least relativistic electrons
and magnetic field, as found in the lobes of radio galaxies.

In the first case, the material would simply be a hot thermal plasma,
and thus, a signature of this gas may be expected in the X-ray
spectrum of the bubble.  To investigate this, {\bf we used a method similar
to that of \citet{2006astro.ph.11210S}}; we extracted a spectrum
for the region defined as the cavity in Section~\ref{spatial}.  We fit
the background subtracted (using a blank-sky background), grouped
spectrum with an absorbed \MEKAL model with a temperature fixed to
that of the gas immediately around the cavity, with the normalisation
free to vary.  This model provides a reasonable fit to the cavity
spectrum, but cannot conclusively rule out the presence of very hot
gas in the cavity.  Thus, this component is frozen in the fit and a
second \MEKAL component is added to model the possible hot phase with
fixed abundance (set to 0.3\Zsol) and temperature, leaving only the
normalisation of this component free to vary.  We vary $kT$ by hand
between the temperature ranges of 5--40~keV, and for each value of
$kT$ we obtain a $3-\sigma$ upper limit on the normalisation and
convert this into the maximum pressure of the hot gas at a given
temperature.  If this pressure is significantly greater than the
pressure of the external gas, it implies that the bubble would not be
stable under this configuration (as there may also be an additional
pressure component from non-thermal sources).  However, if the
pressure in the bubble from the putative hot component is less than
the external pressure, this provides a lower limit on the temperature
of any hot gas component.  By varying the temperature of the second
component until the maximum pressure in the hot component exceeds the
external pressure, we can set a limit on the temperature of any hot
component present in the gas.  It should be emphasised that this is
not a 'fit' to the spectrum in the usual sense as there are
insufficient counts to fit a two-component model.  The method
described merely provides a way of testing for the possible presence
of a second component.

As the bubble is in direct mechanical contact with the IGM at a range
of different radii (due to the fact that it is three dimensional),
using a single temperature from the deprojected temperature profile
would not be an accurate reflection of the temperature around the
cavity.  However, using a range of temperatures would not provide a
satisfactory constraint on the pressure.  We thus obtain the projected
temperature in an annulus at the radius of the centre of the cavity,
to provide a more representative temperature and pressure constraint.
Fitting the third annulus from the centre of the \Chandra data with a
simple absorbed \MEKAL model in the 0.5--5.0~keV energy range and
fixing the abundance to 0.3\Zsol, we obtain an external temperature of
$1.41\pm 0.1$~keV.  Assuming that the lower limit of the pressure of
the gas is approximately $5\times 10^{-12}~\mathrm{dyne\ cm^{-2}}$, an
upper limit for $T_H$ can be found.  We obtain a limit of 10--20~keV
for the temperature of any putative hot component; above this
temperature range, the gas is over-pressured with respect to the
external pressure.  Thus, we cannot rule out the presence of hot
tenuous gas in the bubble.

In the second case, as discussed in Section~\ref{introduction}, the
combination of electron energy spectrum and magnetic field must be
such that the bubble produces no detectable radio or inverse-Compton
X-ray emission. For simplicity we consider a uniform electron spectrum
and magnetic field strength throughout the bubble, and take the
magnetic field to be `tangled' so that it has an isotropic magnetic
stress tensor \citep{Leahy1991}. Let the electron energy spectrum be
$N(E)$ such that $N(E) {\rm d}E$ is the number density of electrons
with energies between $E$ and $E+{\rm d}E$, and let the energy density
in non-radiating particles be some multiple $\kappa$ of the energy
density in electrons, as is conventional: then the requirement of
pressure balance gives us
\begin{equation}
{{B^2}\over{2\mu_0}} + (1+\kappa) \int_{E_{\rm min}}^{E_{\rm max}} E
N(E) {\rm d} E = 3p_{\rm ext}
\label{rpb}
\end{equation}
where $B$ is the magnetic field strength and $E_{\rm{min}}$ and
$E_{\rm{max}}$ are the limits on the electron energy densities to be
considered. If we further assume some physically reasonable form for
the electron energy density, such as a power law in energy ($N(E) =
N_0 E^{-p}$), then, for a given $B$, $\kappa$, $E_{\rm{min}}$ and
$E_{\rm{max}}$ we can solve equation \ref{rpb} for the normalization
$N_0$ of the electron spectrum, and so, effectively, for the total
number or energy density of the electron (particle) population.

Pressure balance has been investigated using this basic formalism for
a variety of different types of {\it filled} cavities, with differing
results. In powerful FRII \citep{fr1974} radio galaxies, it has been
possible to measure the magnetic field strengths in the lobes of some
systems by detection of inverse-Compton scattering of the cosmic
microwave background (CMB): the field strengths turn out to be
comparable to, though slightly below, the equipartition values for an
electron/positron ($\kappa=0$) plasma only
\citep{2005ApJ...626..733C}, so that $B \approx B_{\rm eq}$, where
\begin{equation}
{{B_{\rm eq}^2}\over{2\mu_0}} = (1 + \kappa) \int_{E_{\rm min}}^{E_{\rm max}} E
N(E) {\rm d} E
\label{equip}
\end{equation}
and the corresponding plasma+magnetic field pressures are, in several
cases, comparable to the measured external pressures
\citep{2002ApJ...581..948H,2004MNRAS.353..879C}, suggesting that
$\kappa$ is indeed $\approx 0$ in these systems.  Even where no
external pressure can be measured the fact that the measured magnetic
field is close to the equipartition value for $\kappa = 0$ has to be a
systematic coincidence if $(1+\kappa) \gg 1$ in these systems. On the
other hand it has been known for many years
\citep[e.g.,][]{1988A&A...189...11M,1992A&A...265....9F,1998MNRAS.296.1098H,2000ApJ...530..719W}
that the minimum internal pressure ($\kappa = 0$, $B \approx B_{\rm
  eq}$) in the lobes of low-power FRI radio galaxies often falls
orders of magnitude below the measured external pressure:
\cite{2005MNRAS.364.1343D} have recently commented on this effect in a
sample of cluster-centre objects. In these systems it seems that
either $(1 + \kappa) \gg 1$ (as a change in the filling factor of the
plasma has a similar effect to an increase in $\kappa$, we do not
consider this separately) or there is a strong departure from
equipartition. In some cases the lack of detectable inverse-Compton
emission has been used to argue that it is not possible that $B \ll
B_{\rm eq}$. In the following section we discuss the constraints that
the synchrotron and inverse-Compton limits on the NGC 741 cavity place
on the nature of the cavity-filling plasma.

\subsection{Constraints on the nature of the plasma}
The key free parameters of eq. \ref{rpb} are $B$, $\kappa$,
$E_{\rm{min}}$ and $E_{\rm{max}}$, and $p$. We cannot constrain all of
these from our observations.  However, it seems unlikely that the
plasma is magnetically dominated, and so one constraint that we can
impose is that the magnetic field should be less than or equal to the
equipartition value, $B \la B_{\rm eq}$. The limits on observed
synchrotron and inverse-Compton emissivity place additional
constraints on the parameters of the plasma: synchrotron emissivity at
a given frequency constrains a combination of $N_0$, $B$ and $E_{\rm
max}$, while inverse-Compton emissivity constrains $N_0$ and to some
extent $E_{\rm max}$ only. It is easiest to calculate these
constraints numerically, and we do this using the code of
\citet{1998MNRAS.294..615H}, assuming that the dominant photon field
for inverse-Compton scattering is the cosmic microwave background
radiation (i.e. ignoring the negligible synchrotron self-Compton
component). From the radio data in hand we set an upper limit on the
bubble flux density at 1.4~GHz of 7.4~mJy and at 330~MHz (Jetha et al.
in prep) of 20~mJy. These limits were obtained from VLA observations
of the source, with the 1.4~GHz data being the same as used in
Section~\ref{radio} and the 330~MHz from new observations (Jetha et
al, in prep.). For each observation, since part of the radio source
NGC~742 obscures the cavity, the non-detection limits of the cavity
were taken to be three times the off-source rms noise multiplied by
the square root of the number of beams in the cavity at each
frequency. From the X-ray observations, we impose an upper limit on
the inverse-Compton flux density of 5.8~nJy at 1 keV.

We fix the power-law index $p$ at a relatively flat value (we choose
$p=2.1$ based on the results of \citealt{2005ApJ...626..748Y} for FRI
radio sources) which reflects expectations from particle acceleration
models: although there should almost certainly be a tail of particles
with a steeper value of $p$ due to spectral ageing (as reflected in
the steep spectral index of some observed radio lobe systems) it is
very unlikely that it is correct to extrapolate this spectral index
back to low energies, where the electron loss time-scales are very
long, and the steep-spectrum tail makes a negligible contribution to
the overall energy budget of the electrons.  We conservatively take
the minimum electron energy to be $E_{\rm min} = m_{\rm e}c^2$,
i.e. $\gamma_{\rm min} = 1$, where $\gamma$ is the Lorentz factor of
the electron). For a given value of $\kappa$, we can investigate the
constraints on $E_{\rm max}$ ($\gamma_{\rm max}m_{\rm e}c^2$) and $B$
given by the observations on the assumption of pressure balance as in
equation \ref{rpb}.

\begin{figure*}
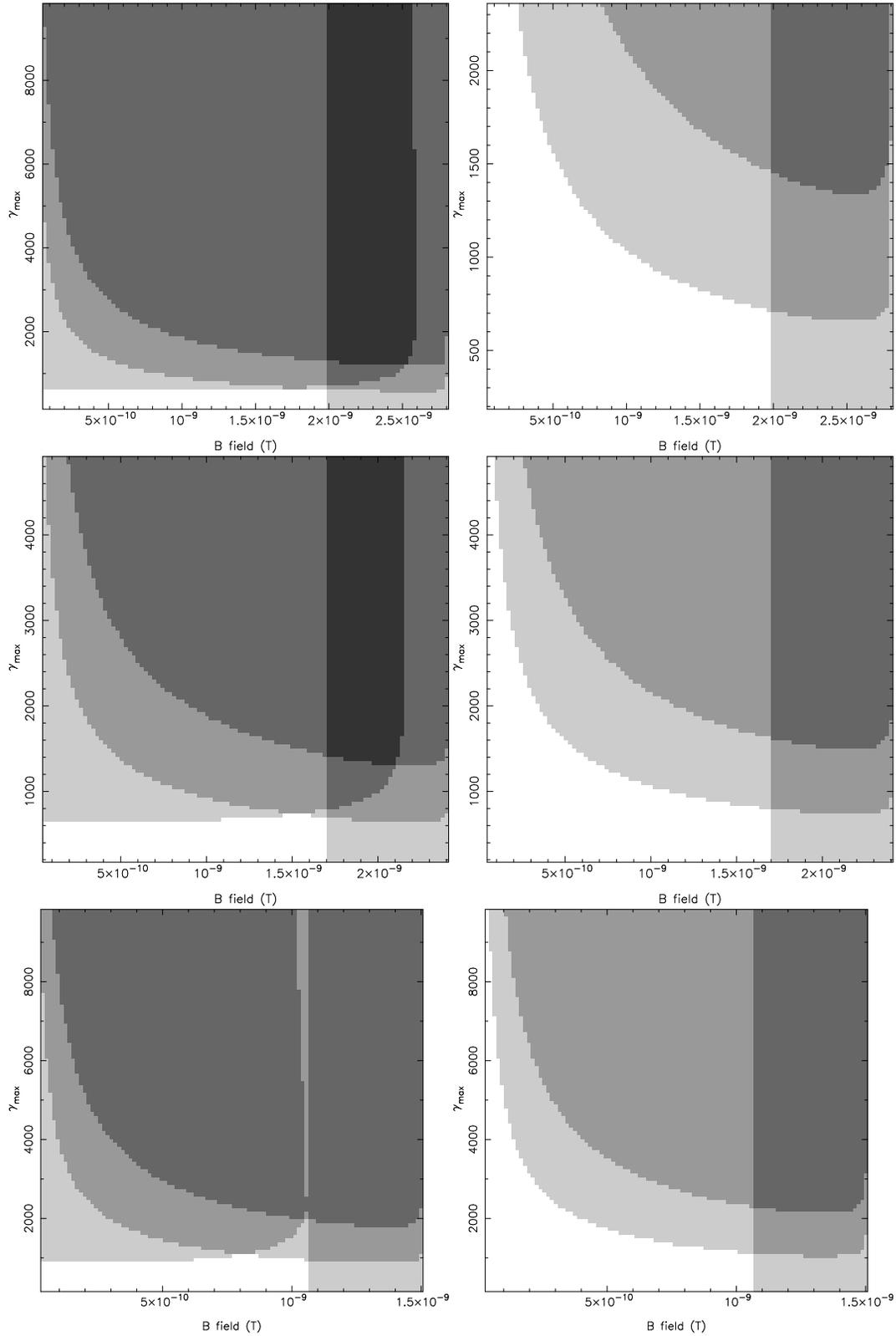

\epsfxsize=7cm
\epsfbox{fig6a.ps}
\epsfxsize=7cm
\epsfbox{fig6b.ps}
\epsfxsize=7cm
\epsfbox{fig6c.ps}
\epsfxsize=7cm
\epsfbox{fig6d.ps}
\epsfxsize=7cm
\epsfbox{fig6e.ps}
\epsfxsize=7cm
\epsfbox{fig6f.ps}
\caption{Constraints imposed by the radio and X-ray limits on the
  physical conditions of any relativistic plasma in the bubble.  The
  white regions indicate those regions of the parameter space allowed
  given the constraints, and the greyscale indicates the number of
  constraints that are violated, ranging from 1 (light grey) to 4
  (both radio limits, the IC limit, and the constraint that the plasma
  should not be magnetically dominated) (dark grey). From top to
  bottom, the panels show the constraints for external pressures of
  $10.4 \times 10^{-12}$, $7.7 \times 10^{-12}$, and $3.0 \times
  10^{-12}$ dyne cm$^{-2}$. The left-hand panels show the $\kappa=0$
  constraints while the right-hand ones show $\kappa=10$. Note the
  different magnetic field and $\gamma_{\rm max}$ scales of the
  three sets of plots.}
\label{constraint-plot}
\end{figure*}

The results are plotted in Fig.~ \ref{constraint-plot} for two
representative values of $\kappa$ as defined above. We chose two
values, $\kappa=0$ (i.e. relativistic electrons/positrons and field
give the only contribution to the internal pressure of the plasma, as
in FRII sources) and $\kappa=10$ (i.e. there is a population of
non-radiating particles, such as protons, with energy density
exceeding that of the electrons by an order of magnitude). We carry
out the calculations using the best estimate of the external pressure
($7.7\times 10^{-12}~\mathrm{dyne\ cm^{-2}}$) and the range imposed by
the uncertainties in temperature, density and cavity position
(3--10.4$\times 10^{-12}~\mathrm{dyne~cm^{-2}}$). It can be seen that
the radio and X-ray data already impose quite severe constraints on
the possible electron energy spectrum if the pressure is close to its
best-fitting value. For $\kappa = 0$ the combination of
inverse-Compton and synchrotron constraints requires $\gamma_{\rm
{max}} < 1000$ for any plausible value of $B$. Even for $\kappa = 10$
the magnetic field strengths must be very low compared to their
equipartition values for $\gamma_{\rm max}$ to be greater than a few
$\times 10^3$: this compares to $\gamma_{\rm {max}} \ga 2 \times 10^4$
for the lobes of typical radio sources with detections at GHz
frequencies (assuming fields close to equipartition). For the values
of $\kappa$ investigated (0 and 10), the bubble can be filled with a
plasma with $\gamma_{\rm max}\ga 2 \times 10^3$ only for the lowest
external pressures {\it and} the lowest internal magnetic fields. It
is clearly possible to choose a value of $\kappa$ such that we have no
constraint on the properties of the electron population -- this
corresponds to a situation in which the energetics of the bubble
plasma are entirely dominated by an `invisible' component. However, it
is interesting to consider the constraints on the electron population
for a situation in which $(1 + \kappa) \approx 1$. For these low
values of $\kappa$, how can we obtain the required low values of
$\gamma_{\rm max}$?

Radiative losses alone (to synchrotron and
inverse-Compton emission) would require a long time to reduce an
initially high $\gamma_{\rm {max}}$ to values $\sim 1000$.
The maximum energy in an electron population that initially extended
to high energies after a time $t$ is given by \citet{1970ranp.book.....P}
\begin{equation}
\gamma_{\rm {max}} = {3\over 4} {{m_{\rm {e}} c}\over{\sigma_{\rm {T}} U t}}
\end{equation}
where $m_e$ is the mass of the electron, $\sigma_T$ is the Thomson
cross-section and $U$ is the total energy density in magnetic field
and microwave background photons (i.e. the total energy density
implicated in loss processes). For equipartition magnetic field
strengths synchrotron losses dominate, $U = 3p_{\rm ext}/2$ and the
age of the bubble is $> 2 \times 10^8$ years (though note that this
calculation takes no account of the fact that the magnetic field may
have been higher at earlier times). For magnetic field strengths much
less than equipartition inverse-Compton losses dominate, $U =
8\pi^5(kT_{\rm CMB})^4/15(hc)^3$, and the limit on the age of the
bubble is closer to $> 2 \times 10^9$ years, where $t$ is the time
since any high-energy particle acceleration took place in the source.
These time-scales are considerably longer than the time-scales for the
bubble to rise buoyantly to its current position ($8.6\times 10^7$
years, Section~\ref{energetics}).  $\gamma_{\rm max}$ may also be
reduced by adiabatic expansion. which gives $\gamma_{\rm max} \propto
V^{-(\Gamma-1)}$, where $\Gamma$ is the adiabatic index, 4/3. To take
a simple example, we could suppose that the original bubble had
$\gamma_{\rm max} = 2 \times 10^4$ as quoted above and has expanded
adiabatically. Then the volume must have changed by a factor 8000
during the expansion; i.e. the initial radius of the bubble was around
1 kpc. But this would require the initial bubble to have been over
pressured by a factor $1.6 \times 10^5$ if it is in pressure balance
now, which seems unlikely. As shown by Alexander (1987), more modest
expansion factors coupled with radiative losses do not greatly change
the inferred ages.

We can therefore conclude that it seems hard for the bubble to
have evolved from a `normal' region of radio-emitting plasma -- like
the plumes of NGC 742 whose radio emission we see in the system at
present -- if the lobe external pressure is close to the best-fitting
value.  The present X-ray data do not allow us to constrain the
pressure around the cavity well enough to rule this possibility out
completely, though, and a deeper X-ray observation is required to
allow us to draw firm conclusions: if the pressure is at the low end
of the range we estimate, and the internal pressure of the cavity is
dominated by non-radiating relativistic protons with a very low
magnetic field, then the available constraints are much weaker. In a
future paper we will apply the technique used above to constrain the
nature of the cavity-filling plasma in other `ghost cavity' systems.

\section{Conclusions}
\label{conclusions}

We have examined the energetics of the galaxy group associated with
NGC~741.  We find that if the cavity seen in the X-ray data was
originally inflated by a radio-loud AGN associated with the brightest
galaxy, NGC~741, then there is sufficient
energy to counteract cooling. This interpretation is consistent with
the fact that NGC~741 clearly has an active nucleus at the present
time, and this could have been more active in the past.

We have investigated the nature of the cavity-filling plasma. We
cannot rule out the possibility that the cavity is filled with a very
hot thermal plasma ($kT > 10$ keV) or with a relativistic plasma that
is energetically completely dominated by non-radiating particles.
However, it is not at all clear that current models for AGN inflation
of cavities can produce cavities of this kind, particularly when we
take into account the relatively small distance (30 kpc) between the
cavity and the centre of the host group. If on the other hand there is
a substantial contribution to the energy density in the cavity from a
plasma that contains relativistic electrons and magnetic field, as in
the active lobes of radio galaxies (such as NGC 742 in the same group)
then we find that the limits placed by the non-detection of radio
synchrotron and X-ray inverse-Compton radiation severely constrain the
maximum electron Lorentz factor. In this scenario, we find that it is
very hard for the cavity to have evolved to its current state from a
`dead' radio lobe by a combination of the standard processes expected
to be operating (synchrotron, inverse-Compton, and adiabatic expansion
losses); for radiative losses only the lobe would have to be older
($\ga 2 \times 10^8$ years) than any plausible bubble inflation time
(between $5 \times 10^7$ and $9 \times 10^7$ years). The extreme
parameters required for the NGC 741 cavity, and for other `ghost
cavities' whose physical parameters have been investigated (e.g. Dunn
et al 2005) and the fact that cavities are found at a wide range of
radii (Birzan et al 2004) may indicate that inflation by the lobes of
a standard radio-loud AGN is not a viable explanation for all ghost
cavities. In the case of NGC 741, we are left with one clearly
outstanding question: if the cavity in NGC 741 is due to an
AGN-inflated bubble, in what form was the energy injected into the
IGM, and in what form is it now?

\section*{Acknowledgements}

NNJ thanks CNES (the French Space Agency) for funding.  MJH thanks the
Royal Society for a research fellowship.  AB is supported by NSERC
(Canada) Discovery Grant program. He is grateful to the Leverhulme
Trust for awarding him a Leverhulme Visiting Professorship, and thanks
University of Oxford, Institute for Computational Cosmology (Durham)
and University of Birmingham for their hospitality during his frequent
visits.  EO'S acknowledges support from NASA grant NNX06AE90G-R.  The
authors would like to thank Monique Arnaud for her help with the \XMM
data reduction and for useful conversations regarding the data, and
Gabriel Pratt and Etienne Pointecouteau for their help with the \XMM
data reduction software used in this paper.  The authors also thank
the anonymous referee for their useful comments in improving the text.
\bibliographystyle{mn2e}
\bibliography{MN-07-1395-MJ_R2}

\label{lastpage}

\end{document}